\documentclass[10pt,twocolumn,letterpaper]{article}

\usepackage{iccv}
\usepackage{times}
\usepackage{epsfig}
\usepackage{graphicx}
\usepackage{threeparttable}
\usepackage{amsmath}
\usepackage{amssymb}
\usepackage{bm}
\usepackage{subcaption}
\usepackage{booktabs}
\usepackage{multirow}
\usepackage{makecell}
\usepackage[accsupp]{axessibility}
\usepackage{cite}
\usepackage{url}

\usepackage{color}

\makeatletter
\DeclareRobustCommand\onedot{\futurelet\@let@token\@onedot}
\def\@onedot{\ifx\@let@token.\else.\null\fi\xspace}

\def\ie{\emph{i.e}\onedot} 
 
 \def\vs{\emph{vs}\onedot}
 
\def\etal{\emph{et al}\onedot}

\makeatother

\newcommand{\sref}[1]{\textcolor{red}{\hyperref[#1]{(\subref*{#1})}}}

\usepackage[pagebackref=true,breaklinks=true,colorlinks,bookmarks=false]{hyperref}

\usepackage[capitalize]{cleveref}
\crefname{section}{Sec.}{Secs.}
\Crefname{section}{Section}{Sections}
\Crefname{table}{Table}{Tables}
\crefname{table}{Tab.}{Tabs.}

 \iccvfinalcopy 

\def\iccvPaperID{7309} 

\begin{document}

\title{RawHDR: High Dynamic Range Image Reconstruction from \\ a Single Raw Image}

	\author{Yunhao Zou\textsuperscript{$1$} \qquad \qquad Chenggang Yan\textsuperscript{$2$} \qquad \qquad  Ying Fu\textsuperscript{$1$}\thanks{Corresponding Author: fuying@bit.edu.cn} \\
		\textsuperscript{$1$}Beijing Institute of Technology \quad\quad \textsuperscript{$2$}Hangzhou Dianzi University \\ 
}

\maketitle

\begin{abstract}
High dynamic range (HDR) images capture much more intensity levels than standard ones. Current methods predominantly generate HDR images from 8-bit low dynamic range (LDR) sRGB images that have been degraded by the camera processing pipeline. However, it becomes a formidable task to retrieve extremely high dynamic range scenes from such limited bit-depth data. Unlike existing methods, the core idea of this work is to incorporate more informative Raw sensor data to generate HDR images, aiming to recover scene information in hard regions (the darkest and brightest areas of an HDR scene). To this end, we propose a model tailor-made for Raw images, harnessing the unique features of Raw data to facilitate the Raw-to-HDR mapping. Specifically, we learn exposure masks to separate the hard and easy regions of a high dynamic scene. Then, we introduce two important guidances, dual intensity guidance, which guides less informative channels with more informative ones, and global spatial guidance, which extrapolates scene specifics over an extended spatial domain. To verify our Raw-to-HDR approach, we collect a large Raw/HDR paired dataset for both training and testing. Our empirical evaluations validate the superiority of the proposed Raw-to-HDR reconstruction model, as well as our newly captured dataset in the experiments.
\end{abstract}


\section{Introduction}
\label{sec:intro}
The dynamic range of real-world scenes often surpasses the recording capability of standard consumer camera sensors, leading images to lose details in both over- and under-exposed regions~\cite{battiato2003high}. To endow today's digital photos with the capacity to contain more scene information, the technique called high dynamic range (HDR) that records data with a wide range of intensity levels has been extensively explored in the computational imaging community~\cite{wang2021deep}.  Compared with conventional low dynamic range (LDR) images, HDR retains more details in over- and under-exposure regions. Thus, HDR benefits downstream vision tasks including segmentation~\cite{martinez2017image}, object detection~\cite{onzon2021neural}, and also provide more aesthetically appealing pictures~\cite{hasinoff2016burst, kim2019deep}, which computer vision researchers have longly pursued.

\begin{figure}[t]
	\centering
	\begin{subfigure}[b]{.3\linewidth}
		\centering
		\includegraphics[width=\linewidth]{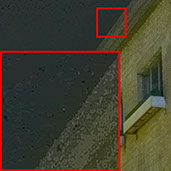}
		\vspace{-6mm}
		\subcaption*{Dark (RGB)}
	\end{subfigure}
	\begin{subfigure}[b]{.3\linewidth}
		\centering
		\includegraphics[width=\linewidth]{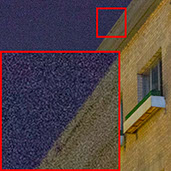}
		\vspace{-6mm}
		\subcaption*{Dark (Raw)}
	\end{subfigure}
	\begin{subfigure}[b]{.3\linewidth}
		\centering
		\includegraphics[width=\linewidth]{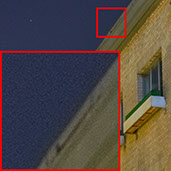}
		\vspace{-6mm}
		\subcaption*{Dark (HDR)}
	\end{subfigure}
	\begin{subfigure}[b]{.3\linewidth}
		\centering
		\includegraphics[width=\linewidth]{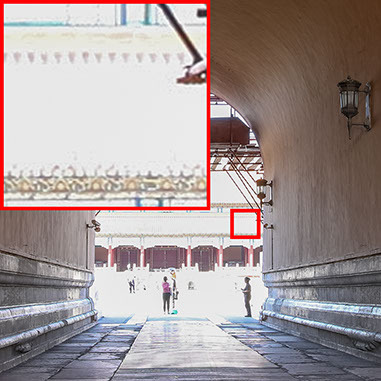}
		\vspace{-6mm}
		\subcaption*{Bright (RGB)}
	\end{subfigure}
	\begin{subfigure}[b]{.3\linewidth}
		\centering
		\includegraphics[width=\linewidth]{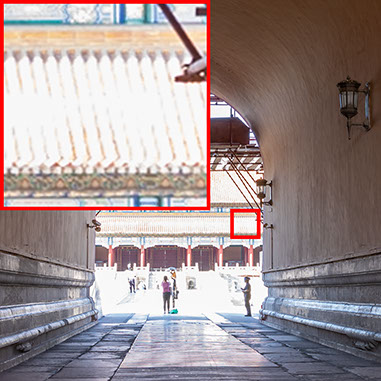}
		\vspace{-6mm}
		\subcaption*{Bright (Raw)}
	\end{subfigure}
	\begin{subfigure}[b]{.3\linewidth}
		\centering
		\includegraphics[width=\linewidth]{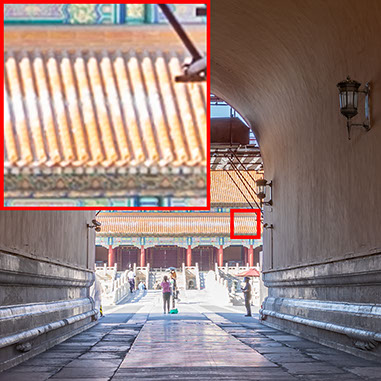}
		\vspace{-6mm}
		\subcaption*{Bright (HDR)}
	\end{subfigure}
	\vspace{-1mm}
	\caption{The RGB/Raw/HDR images of the darkest (first row) and brightest (second row) regions in a high dynamic scene. This paper is motivated by two observations: (1) HDR scenes contain both extremely dark and bright regions, which are very challenging to reconstruct from a single image; (2) Raw images contain much more information in these hard regions, compared to low-bit RGB images.}
	\label{fig:datacompare}
	\vspace{-3mm}
\end{figure}

Methods to obtain HDR data broadly fall into three categories, \ie, reconstruction from multi-exposure, single-exposure images, and novel camera sensors. Among new sensors, some noteworthy examples include HDR cameras~\cite{robidoux2021end, metzler2020deep}, event cameras~\cite{wang2019event,rebecq2019high,han2020neuromorphic,zou2021learning}, and infrared sensors~\cite{liu2020deep}. Since these sensors are all specialized devices, more works focus on the reconstruction from multi-/single-exposures captured by commercial cameras, which are more practical and commercially friendly.

Historically, considering that single exposure cannot record the intensity information of a scene that covers a large dynamic range, researchers often combined multi-exposure images to generate HDR content~\cite{hasinoff2016burst, kalantari2017deep,choi2020pyramid,yan2019multi,yan2019attention,endo2017deep}. The quality of these images greatly depend on aligning different exposures, and may suffer from ghosting effect caused by imperfect alignment.

To avoid the potential risk of alignment failure, more recent works incorporate only a single LDR image to reconstruct an HDR image~\cite{eilertsen2017hdr,liu2020single,xing2021invertible,chen2021hdrunet}. Nonetheless, single-image HDR reconstruction is more challenging due to the physical limitation of consumer camera's dynamic range. Consequently, under-exposed regions are often noisy~\cite{chen2018learning,chen2023instance, fu2022gan, zhang2021learning}, while over-exposed regions are difficult to recover~\cite{afifi2021learning, huang2022mm}.  Most previous works utilize low-bit sRGB images for HDR reconstruction. However, sRGB images have been degraded by lossy in-camera operations, which is not enough to record details in an HDR scene. Even as innovative and interpretable deep models emerge~\cite{eilertsen2017hdr,liu2020single}, the HDR reconstructions from single LDR images often falter in highly dynamic scenes due to the inherent paucity of low-bit sRGB input.

In this work, we aim to relieve the information limitation for single-exposure setting with a specialized reconstruction model and high-quality dataset.  Using a single image, the challenge comes down to the recovery of the darkest and brightest regions (\textit{hard} regions) in high dynamic scenes. The visualization in Figure~\ref{fig:datacompare} shows that commonly used sRGB images contain limited  information in hard regions. Raw images retain more details than sRGB, but are still far from HDR. Therefore, we propose using unprocessed Raw sensor data, which has higher available bit-depth and better intensity tolerance, thus can circumvent the long-standing drawback of insufficient scene information. To perform specific operations on hard regions, we learn an exposure mask to adaptively separate the over-/under- and well-illuminated regions for each scene. We also devise a deep neural network specially designed for Raw input images to exploit the information in hard regions. Crucially, we propose a dual intensity guidance based on the channel-variant attribute of Raw images to guide less informative image channels and global spatial guidance to well exploit longer-range information. Finally, we collect a high-quality Raw/HDR paired dataset for both training and testing. The quality of our Raw-to-HDR reconstruction methods and dataset are verified in the experiments. 

Our main contributions can be summarized as follows:
\begin{enumerate}
	\item We focus on the essential issue of HDR imaging --- the challenge in recovering the dark and bright regions, for which  we propose to learn an exposure mask to separate the image into hard and easy regions.
	\item  We propose a deep network to deal with the hard regions, including a dual intensity guidance built on the channel-variant attribute of Raw images and a global spatial guidance built on transformer with spatial attention that exploits information from a longer range.
	\item We directly reconstruct HDR from a single Raw image, which is endowed with higher bit-depth to handle high dynamic scenes and can be potentially integrated into modern camera processing pipelines. In addition, we collect a high-quality paired Raw/HDR dataset for training and evaluation.
\end{enumerate}
\section{Related Work}
In this section, we review the most relevant works, including multi-image HDR reconstruction and single-image HDR reconstruction.

\vspace{1mm}
\noindent\textbf{Multi-image HDR reconstruction.} In early years, a number of researchers reconstruct HDR images by merging a series of bracketed exposure LDR images~\cite{debevec2008recovering}. For multi-exposure HDR imaging, the step to align multiple images is crucial yet challenging. Numerous   methods~\cite{kalantari2017deep,yan2019multi,peng2018deep} align bracket multi-exposures before fusing. Given a sequence of LDR images with different exposures, Kalantari \etal~\cite{kalantari2017deep} regarded the medium exposure as reference and align low- and high-exposure images to the reference by flow warping. Then they learned the mapping from aligned LDR to HDR through a deep neural network. 
Peng \etal~\cite{peng2018deep} investigated the potential of advanced optical flow estimation techniques, such as FlowNet~\cite{dosovitskiy2015flownet}, to refine alignment before HDR reconstruction. In addition to fusing different exposures, HDR+~\cite{hasinoff2016burst} fused burst images under the same exposure and claimed that it was easier to align images with the same exposure. However, all these methods would encounter difficulties in alignment in very high dynamic scenes. Therefore, multi-exposure HDR reconstruction is mainly used in situations without significant motion.

\vspace{1mm}
\noindent\textbf{Single-image HDR reconstruction}. Given the inherent challenges with multi-exposure HDR, especially the difficulty in LDR alignment, some researchers have gravitated towards deriving HDR from a single image. However, the recording capacity of a single image makes it even more ill-posed. Traditional single-exposure HDR works proposed to expand the dynamic range by estimating the density of light sources ~\cite{banterle2009high,banterle2007framework}. As recent deep learning-based tools, especially CNNs, have made a great breakthrough in computer vision, CNN-based methods~\cite{he2016deep, liu2020single} were presented to directly learn to reconstruct HDR from a single LDR image. HDRCNN~\cite{eilertsen2017hdr} and ExpandNet~\cite{marnerides2018expandnet} proposed to directly learn the mapping from LDR-to-HDR  in an end-to-end manner. HDRUNet~\cite{chen2021hdrunet} and SingHDR~\cite{liu2020single} designed the network architecture following the physics formation model of LDR images. Other methods~\cite{lee2018deep} first synthesize pseudo-multi-exposure from a single exposure, then fuse these generated multi-exposures. All of these single-exposure HDR methods~\cite{banterle2009high,banterle2007framework, he2016deep, liu2020single, eilertsen2017hdr, lee2018deep} tried to improve HDR quality from the perspective of designing a better reconstruction algorithm. However,  using sRGB LDR images as input prevents them from obtaining higher-quality HDR. In fact, during the intricate processing pipeline to generate sRGB, there are lossy and invertible in-camera operations including nonlinearization, clipping, compression and quantization, which degrade the original Raw images (generally 14-bit) to lower bit-depth images (generally 8-bit). Therefore, we believe that the linearity and high bit-depth attributes of untouched Raw images are ideal for single-image HDR reconstruction. In this paper, we reconstruct HDR images directly from a single Raw image, with customized network design and specially captured paired dataset.



\section{Method}
This section demonstrate our primary motivations, problem formulation for Raw-to-HDR reconstruction, and the network architecture of our proposed RawHDR model. The overarching methodology is depicted in Figure~\ref{fig:overview}.

\subsection{Motivation}
\label{sec:motivation}

Existing sing-exposure HDR reconstruction methods~\cite{liu2020single,chen2021hdrunet,eilertsen2017hdr} mainly focus on the reconstruction from low-bit sRGB images. Despite their ability to improve the quality beyond the original sRGB images, they cannot handle hard regions in extremely high dynamic scenes. This limitation lies in the irreversible and lossy operations in the in-camera signal processing workflow, including nonlinearization, clipping, compression and quantization~\cite{wei2021physics,karaimer2016software,zou2022estimating}. Furthermore, these methods target at enhancing previously captured LDR images rather than optimizing the image capture process itself. Motivated by these shortcomings, we plan to reconstruct HDR images in higher dynamic scenes with a novel reconstruction method and new data setting. To resolve the problem of information loss and insufficient details in commonly used sRGB images, we directly utilize the unprocessed Raw data to reconstruct HDR images, as Raw images have higher bit-depth to maintain a large dynamic range of a scene. Moreover, since we directly perform HDR mapping on Raw images that are unprocessed, the operation can be potentially integrated into today's camera processing pipeline to facilitate the imaging process. 

\begin{figure}[t]
	\centering
	\small
	\begin{subfigure}[b]{.45\linewidth}
		\centering
		\includegraphics[width=\linewidth]{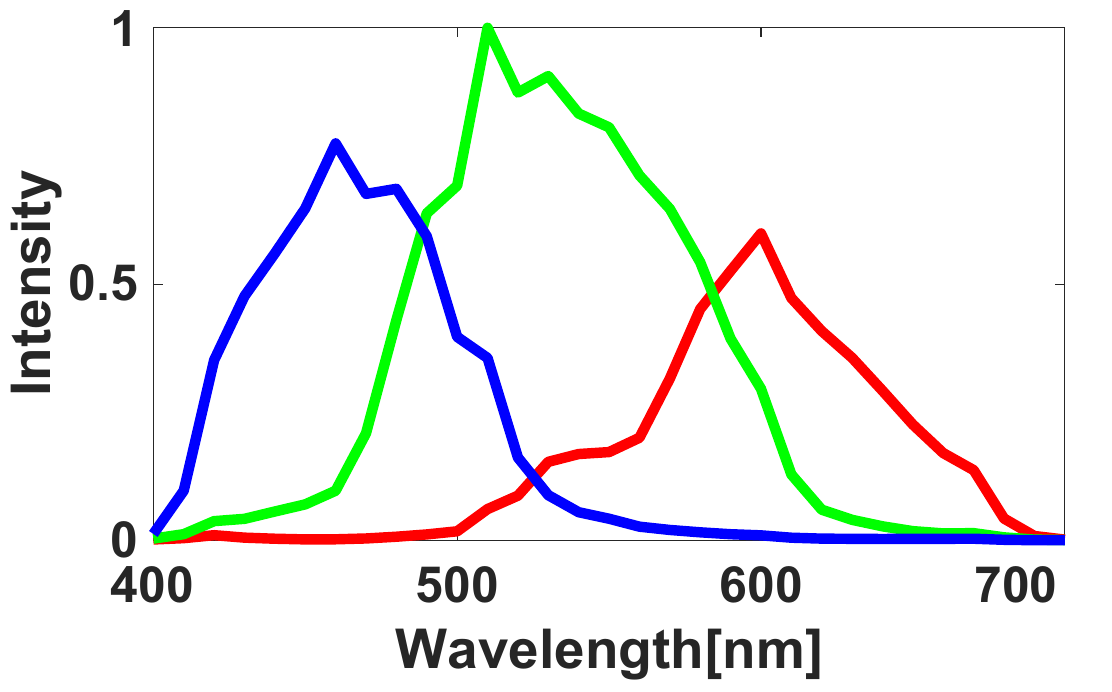}
		\vspace{-5mm}
		\caption{CRF}
		\label{fig:crf}
	\end{subfigure}
	\hspace{4mm}
	\begin{subfigure}[b]{.43\linewidth}
		\footnotesize
		\setlength{\tabcolsep}{3mm}
		\centering
		\begin{tabular}{c|c}
			\toprule
			Channel & Mean value\\
			\midrule
			Red & 704.93 \\
			\midrule
			Green & 1273.61\\
			\midrule
			Blue & 942.00 \\
			\bottomrule
		\end{tabular}
		\vspace{0.5mm}
		\caption{Channel Mean}
		\label{fig:channel}
	\end{subfigure}
	\\[1mm]
	\begin{subfigure}[b]{\linewidth}
		\centering 
		\footnotesize
		\setlength{\tabcolsep}{1mm} 
		\begin{tabular}{cccc}
			\hspace{-1mm}
			\includegraphics[width=.24\linewidth,clip,keepaspectratio]{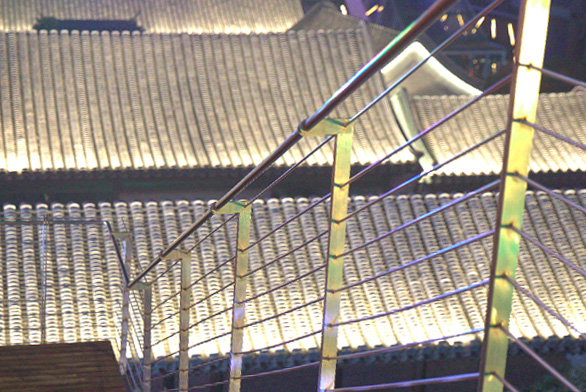} &\hspace{-2mm}
			\includegraphics[width=.24\linewidth,clip,keepaspectratio]{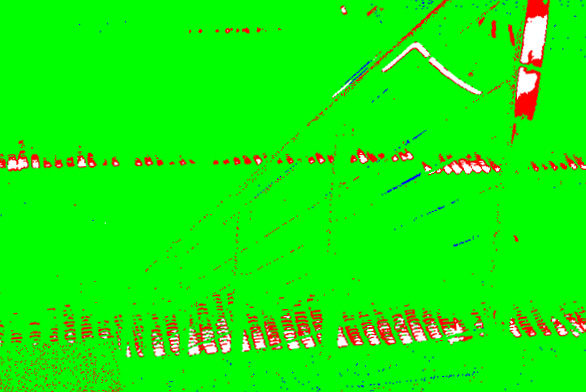} &\hspace{-2mm}
			\includegraphics[width=.24\linewidth,clip,keepaspectratio]{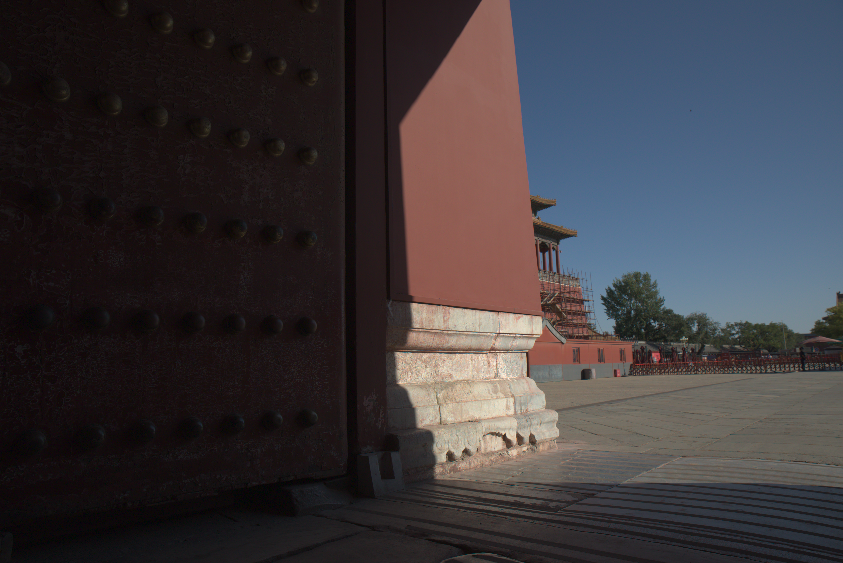} & \hspace{-2mm}
			\includegraphics[width=.24\linewidth,clip,keepaspectratio]{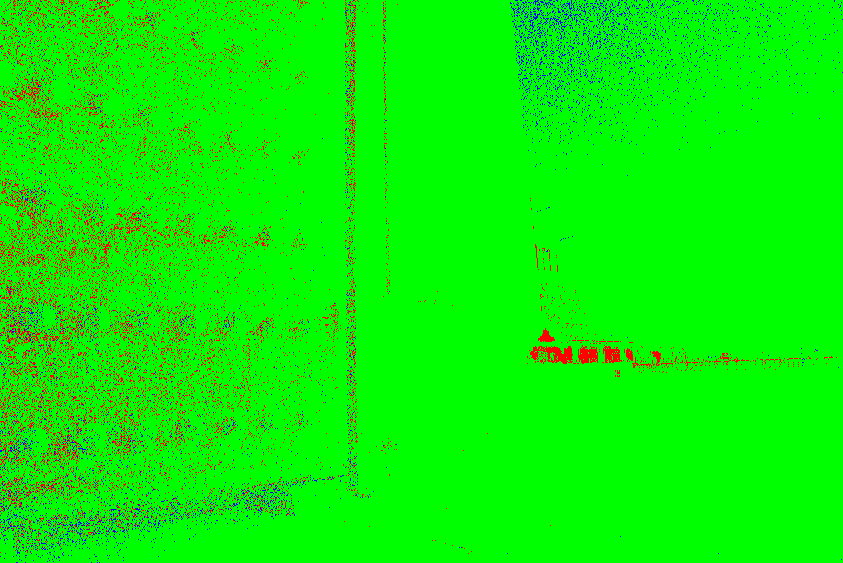} \\
			Scene 1 & DCR 1 & Scene 2 & DCR 2 \\	
		\end{tabular}
		\vspace{-2mm}
		\caption{Examples for dominant color channel (denoted as DCR).}
		\label{fig:dcr}
	\end{subfigure}
	
	\vspace{-2mm}
	\caption{Analysis for the channel-wise attributes of Raw images. (a) A typical camera response function; (b) The channel-wise mean values for the captured dataset; (c) Examples of two scenes, with the original captured images and  the corresponding pixel-wise dominant color channels.}
	\label{fig:statics}
	\vspace{-2mm}
	
\end{figure}

Then, we carefully analyze the special attribute of Raw images to present a deep model tailored for Raw-to-HDR mapping. Notably, the intensity values vary with image channels in Raw space~\cite{liu2020joint}. We provide the camera response function (CRF) of a typical consumer camera (Canon 5D II) in Figure~\ref{fig:statics}\sref{fig:crf}. It illustrates that the integration of green spectral curve is considerably larger than red and blue, which indicates higher sensitivity of green patterns. Consequently, green channels have larger values in Raw images. This conclusion is also verified in Figure~\ref{fig:statics}\sref{fig:channel}, where the channel-wise mean values for our dataset (described in Section~\ref{sec:dataset}) is presented. Actually, apart from regions with a strong inclination towards red or blue, the pixel values in most areas tend to be higher for green. We give example scenes in Figure~\ref{fig:statics}\sref{fig:dcr}, which shows the predominant channel for each pixel, we can see that even the \textit{blue sky} and \textit{red wall} have larger green value in RAW space.

Therefore, green channels of Raw images are more likely to lose information due to the intensity upper bound.  Red and blue channels face similar situations in poorly lit areas.  Utilizing this attribute, we lean on the more informative channels to guide the others. Moreover, considering the severe information loss in hard areas (Figure~\ref{fig:datacompare}), we introduce longer range feature exploitation models for the reconstruction of hard regions to compensate for insufficient information in these regions. In addition, since we plan the handle hard and easy regions separately, we propose to learn exposure masks to achieve this task.

\subsection{Model Architecture}

\begin{figure*}[t]
	\centering
	\includegraphics[width=0.85\linewidth, clip, keepaspectratio
	]{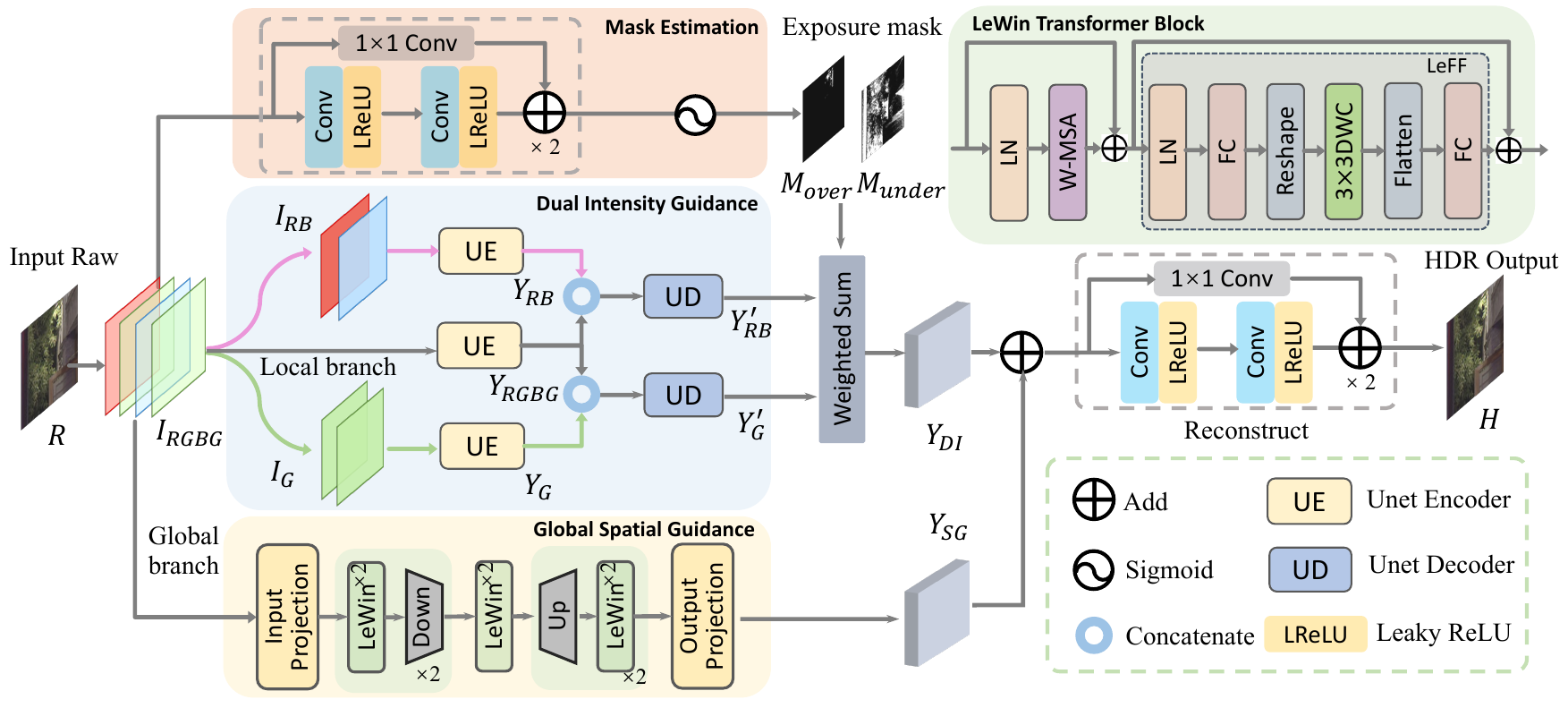}
	\vspace{-2.5mm}
	\caption{The overview of our HDR reconstruction method with dual intensity and global spatial guidance.}
	\vspace{-2mm}
	\label{fig:overview}
\end{figure*}

Given a Raw image $R$ and a deep neural network model $f(\cdot, \theta)$, the process to reconstruct an HDR image $H$ can be formulated as
\begin{equation}
	H = f(R, \theta),
\end{equation}
where $\theta$ is the network parameters. 
The overview of our RawHDR model is depicted in Figure~\ref{fig:overview}. We first learn exposure masks through a mask estimation module, which separate the over-/under- and well-exposed areas. Then, on the basis of the intensity attribute of Raw images, we present dual intensity guidance to guide less informative channels with more informative ones. Finally, we use a transformer-based model to harness longer-range spatial features, which we call global spatial guidance.

\vspace{1mm}
\noindent\textbf{Mask estimation.}
As we know,  HDR images contain information details in both brightest and darkest areas, which cannot be well captured by ordinary LDR images. To employ customized operations on these regions, we need to find out the mask to separate the over-exposed, under-exposed, and well-illuminated regions of a scene. In this work, we use learnable masks to separate different regions, then perform different operations for these regions.

Previous works prefer to use manual-set thresholds to obtain the masks. For example, \cite{liu2020single} classifies intensities surpassing 0.95 as over-exposed and those below 0.05 as under-exposed, subsequently generating a blending mask from these thresholds. Such masks, acquired through rigid thresholding, are denoted as $M_{over}^h$ and $M_{under}^h$.
However, such mask is too hard to separate the regions, potentially leading to undesirable edge artifacts in smooth areas. Therefore, we propose a gentler and more adaptive thresholding strategy. Here, we use deep neural networks to predict the exposure mask. Our mask estimation model incorporates local feature extraction, and the pixel-wise intensity value is not the only factor in deciding the exposure masks. The process to estimate the masks for over-, under-, and well-exposed regions (denoted as $M_{over}$, $M_{under}$ and $M_{well}$) can be formulated as 
\begin{equation}
	\begin{split}
		&M_{over}=\mathcal S(P_o(R)), \\
		&M_{under}=\mathcal S(P_u(R)),\\
		&M_{well}=\max\{1-M_{over}-M_{under}, 0\},
	\end{split}
\end{equation}
where $\mathcal S$ is the sigmoid function and $P_o$ and $P_u$ are networks to predict the over and under-exposure masks. In our experiment, we simply use two residual blocks~\cite{he2016deep} to implement $P_o$ and $P_u$. In order to guarantee that $M_{over}$ is close to $1$ at high intensity values while $M_{under}$ is close to $1$ at low intensity values, we design a constraint loss $\mathcal L_{mask}$ which lead the mask to for mask learning
\begin{equation}
	\mathcal L_{mask} = \Vert M_{over} - M_{over}^h\Vert_1 + \Vert M_{under} - M_{under}^h\Vert_1.
\end{equation}

\noindent\textbf{Dual intensity guidance.}
Modern cameras typically employ the Bayer pattern for image capture, where green sensors on the CMOS receive a greater luminous flux relative to other color sensors~\cite{Zhang_Fu_Li_2022}. As a result, Raw images have higher intensity values in green channels compared to red and blue. Another notable observation is that different channels share similar edges~\cite{liu2020joint}. Based on the two observations, existing works have already utilized this attribute of sensors by guiding red and blue channels with green channels in applications including denoising and demosaicing~\cite{liu2020joint,guo2021joint,Zhang_Fu_Li_2022}. In this work, we present a \textit{dual intensity guidance} module to recover missing details in hard regions of a high dynamic scene. Since green channels are more sensitive to light (Figure~\ref{fig:statics}), it has higher signal-to-noise ratio (SNR) in under-exposed regions. With higher SNR, green channel can guide other channels during reconstruction since all channels share the same edges. In addition, in our HDR reconstruction case, over-exposure regions also need guidance, as these regions lose much information due to intensity clipping. In these regions, less sensitive channels, \ie, red and blue, have smaller intensity values and are less likely to overexpose. So we use red and blue channels as guidance in over-exposed regions.

The architecture of our dual intensity guidance module is illustrated in Figure~\ref{fig:overview}. Our method first packs Raw images with size $H\times W$ into four channels representing RGBG pixels in Bayer pattern, and obtains a $H/2 \times W/2 \times 4$ input image $I_{RGBG}$. Then, we extract the two green channels as under-exposure guiding image $I_G$, and the red/blue channels served as over-exposure guiding image, denoted as $I_{RB}$. Together with the packed input image $I_{RGBG}$, these three tensors are fed into  U-net~\cite{ronneberger2015u} encoders (UE) to separately obtain hidden features  $Y_G$, $Y_{RB}$, and $Y_{RGBG}$. According to the analysis in Section \ref{sec:motivation},  $I_G$ and $I_{RB}$ are less likely to saturate in under- and over-exposed regions. So we regard the features extracted by these two branches as guidance and concatenate them with $Y_{RGBG}$ to incorporate more details in the darkest and brightest regions. Later, the concatenated tensors are decoded by a UNet decoder (UD), and the output of dual intensity guidance module is obtained by weighted summation of over and under-exposure guided features through an intensity mask. More concretely, we operate under-exposure mask $M_{under}$ on the green channel guided feature and mask the red/blue channel guided feature with over-eposure mask $M_{over}$. The process can be formulated as
\begin{equation}
	\begin{split}
		&Y_G'=D(\text{Concat}(Y_{G}, Y_{RGBG})),\\
		&Y_{RB}'=D(\text{Concat}(Y_{RB}, Y_{RGBG})),\\
		&Y_{DI}=M_{under}\odot Y_G' + M_{over}\odot Y_{RB}',
	\end{split}
\end{equation}
where $\odot$ denotes the Hadamard product, $Y_G'$ and $Y_{RB}'$ are the intensity guided features. 

\vspace{1mm}
\noindent\textbf{Global spatial guidance}.
Our dual intensity guidance module is designed from the perspective of channel-wise guidance. In other words, we guide some channels with more informative channels pixel-wisely. However, in more extreme regions, the most informative channel may still be far from sufficient details. Under these conditions, longer range features can be utilized since there may exist similar patches that can help the recovery of hard regions~\cite{xu2022snr}. Therefore, besides channel-wise local guidance, we further present a spatial guidance branch that exploits long range features. Specifically, we use a series of transformer blocks to extract features, which we call global spatial guidance.

We incorporate a U-net like transformer structure~\cite{wang2022uformer} to build the global spatial guidance. As shown in Figure~\ref{fig:overview}, the global spatial guidance consists of $K$ stages, and each stage contains 2 Locally-enhanced Window (LeWin) transformer blocks~\cite{wang2022uformer}. Among stages, there are down- and up-sampling operations to obtain a large receptive field. The LeWin transformer block is built with attention operations to exploit the relationship between pixels. Based on the U-shape structure and attention mechanism, the global spatial guidance would find similar patches for hard regions from a global receptive field. These additional spatial features introduce more available details and facilitate the reconstruction of hard regions. In addition, the LeWin transformer blocks divide the feature maps into non-overlapping windows~\cite{liu2021swin} and employ attention mechanism within each window. Though the attention for each block only focuses on a single window, the module still reaches an almost global receptive field due to the U-shape structure.

Figure~\ref{fig:overview} illustrates the structure of each LeWin block. For the $i$-th block, given the input feature $F_{i-1}$, the output feature $F_i$ can be obtained by
\begin{equation}
	\begin{split}	
		&F_i'=\text{W-MSA}(\text{LN}(F_{i-1}))+F_{i-1},\\
		&F_i=\text{LeFF}(\text{LN}(F_i'))+F_i',
	\end{split}
\end{equation}
where W-MSA denotes the window-based multi-head attention module~\cite{wang2022uformer}, LeFF is the locally-enhanced feed-forward network, and LN is the layer normalization~\cite{ba2016layer}. In LeFF, tokens are first processed by a linear projection, then reshaped to 2D feature maps. Next, a $3\times 3$ depth-wise convolution layer is applied to extract spatial features, and the results are flattened back to tokens.

\subsection{Learning Details}
Considering that HDR output images have high bit-depth, normal $L_1$ or $L_2$ loss tends to be dominated by bright areas with extensive intensity values. To counteract this, we compute loss functions in the $\log$ space. Given the reconstructed HDR image $H_t$ and the corresponding ground truth $\hat H_t$, we employ $L_2$ loss in the $\log$ space ~\cite{liu2020single} to evaluate the reconstruction fidelity
\begin{equation}
	\mathcal L_{rec} = \sum_{i=1}^{T}\Vert \log H_i-\log \hat {H}_i\Vert^2_2,
\end{equation}
where $T$ is the total number of training samples.

In addition to pixel-level metrics, we also adopt the Learned Perceptual Image Patch Similarity (LPIPS) loss \cite{zhang2018unreasonable} to ensure high-level structural similarity, denoted as $\mathcal L_{LPIPS}$. Incorporating this loss with the mask constraint, the complete loss function for  becomes:
\begin{equation}
	\mathcal L=\mathcal L_{rec} + \tau_1\mathcal L_{LPIPS} + \tau_2\mathcal L_{mask}.
\end{equation}

For our experiments, we empirically set the weights $\tau_1$, $\tau_2$ to $0.5$ each. During the training phase, we initialize our network using Kaiming initialization~\cite{he2015delving}, and minimize the loss using the adaptive moment estimation method~\cite{kingma2014adam}, with a momentum parameter to 0.9. The initial learning rate is $10^{-4}$, which is subsequently divided by $10$ at $1000$-th epoch. Training is conducted with a batch size of $1$ across $2000$ epochs, using the PyTorch deep learning framework\cite{paszke2019pytorch} on an NVIDIA Geforce 3090 GPU.
\section{The Proposed Raw-to-HDR Dataset}
\label{sec:dataset}

\begin{table*}[t]
	\centering
	\setlength{\tabcolsep}{1mm}
	\setlength{\abovecaptionskip}{1.3mm} 
	\small
	\begin{threeparttable}
		\caption{Summary of existing HDR dataset and the proposed Raw-to-HDR dataset.}
		\vspace{-2mm}
		\begin{tabular}{c|c|c|c|c|c|c|c}
			\hline\hline
			Dataset & Year &  Size & Resolution & Real/Syn.  & Input & Output & Application \\
			\hline
			Froehlich \etal\cite{froehlich2014creating} & 2014 & 15 & 1920$\times$1080 & Real &None&12-bit video & HDR video sequences\\
			\hline
			HDREye \cite{nemoto2015visual} & 2015 & 46 & 1920 $\times$ 1080 & Syn. &  8-bit sRGB &16-bit HDR&Static indoor and outdoor scenes\\
			\hline
			Kalantari \etal\cite{kalantari2017deep} & 2017 & 74  & 1500$\times$1000 & Real  & 14-bit Raw & 16-bit HDR & Multi-exposure HDR\\ 
			\hline
			Liu~\etal\cite{liu2020single}  & 2020 & N/A & 1536$\times$1024 & Real+Syn. & 8-bit sRGB & 12-16-bit HDR  & Single-exposure sRGB-to-HDR\\
			\hline
			NITIRE 2021~\cite{perez2021ntire} & 2021 & 1761 &1920$\times$1080& Syn. & 8-bit sRGB & 12-bit HDR & Curated from video sequences\\
			\hline
			Ours &2023 & 324 & 6720$\times$4480& Real & 14-bit Raw & 20-bit HDR & Single-exposure Raw-to-HDR\\
			\hline\hline
		\end{tabular}
		\label{tab:dataset}
	\end{threeparttable}
		\vspace{-2mm}
\end{table*}

Dataset is significant for current data-driven HDR imaging methods. To train the Raw-to-HDR model, we need a high-quality paired dataset to feed the deep neural networks. Existing datasets mainly contain HDR images or paired sRGB/HDR images~\cite{kalantari2017deep, eilertsen2017hdr,froehlich2014creating,nemoto2015visual, perez2021ntire,liu2020single}.  Table~\ref{tab:dataset} illustrates some recent and representative HDR datasets~\cite{wang2021deep}. To the best of our knowledge, there is no high-quality paired Raw-to-HDR dataset that is designed for supervising single-exposure Raw-to-HDR mapping. Therefore, we build a new Raw-to-HDR paired dataset. We first carefully choose HDR scenes. Then, we use the bracket exposure mode to capture images. We fix the camera on a tripod to ensure that there is no vibration during the capture. Then,  three exposure values are used to capture, including -3EV, 0EV, and +3EV. As a result, for each scene, we have three Raw images with varied exposures. Then, the Raw images at 0EV are served  as input images, and we follow the well-known HDR merging method~\cite{debevec2008recovering} to fuse HDR from raw image series, which are served as ground truth. In total, we collect 324 pairs of Raw/HDR images using Canon 5D Mark IV camera. For each scene, images are with a high resolution of $4480\times 6720$, and the final dataset is carefully checked and filtered to exclude misaligned pairs. The input Raw images of our dataset are recorded in 14-bit Raw format, and the corresponding HDR images are 20-bit, with additional image profiles (white balance, color correction matrix) recorded in the file. The dataset is available at \href{https://github.com/jackzou233/RawHDR}{https://github.com/jackzou233/RawHDR}.

\begin{figure}[t] \small
	\centering
	\setlength{\tabcolsep}{1pt}
	\begin{tabular}{cccccc}
		\rotatebox{90}{\quad\;-3EV} &
		\includegraphics[width=.24\linewidth,clip,keepaspectratio]{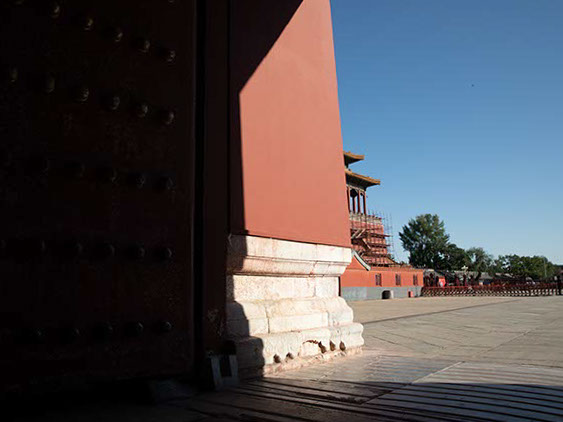} & \includegraphics[width=.24\linewidth,clip,keepaspectratio]{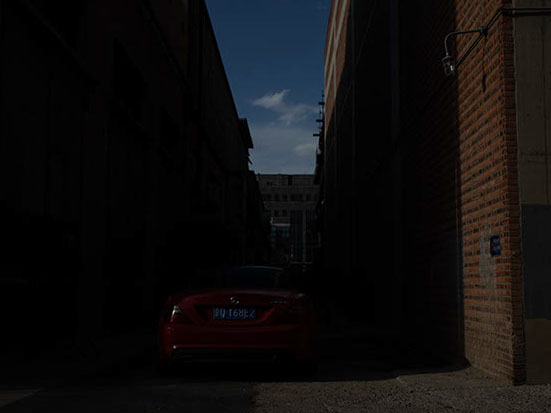} & \includegraphics[width=.24\linewidth,clip,keepaspectratio]{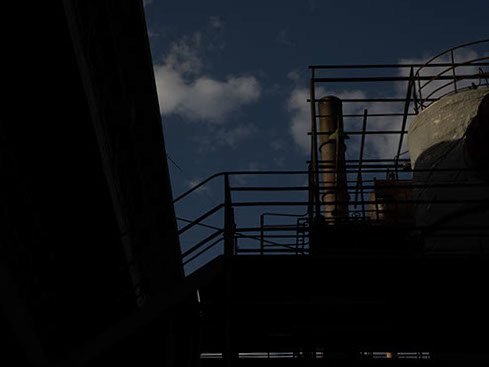} & \includegraphics[width=.24\linewidth,clip,keepaspectratio]{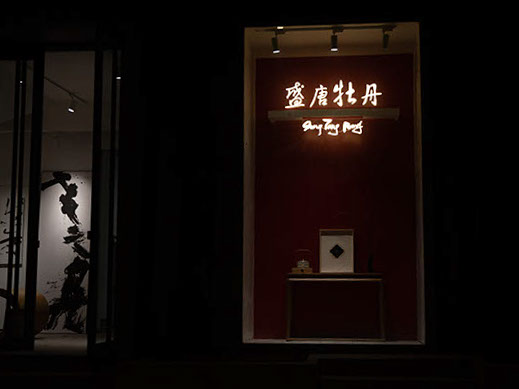} \\
		\rotatebox{90}{\quad\;\;0EV} &
		\includegraphics[width=.24\linewidth,clip,keepaspectratio]{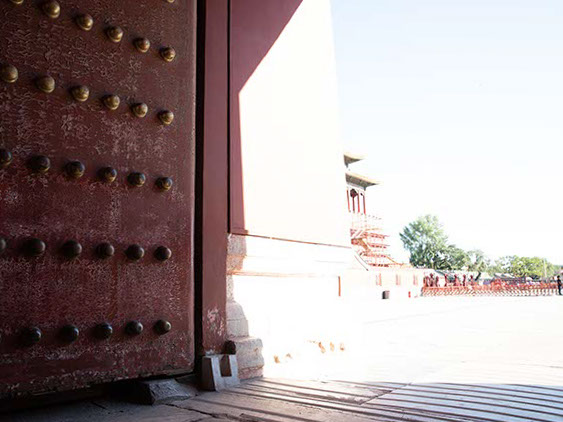} & \includegraphics[width=.24\linewidth,clip,keepaspectratio]{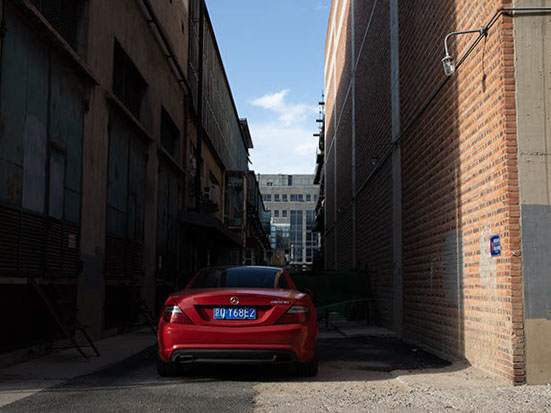} & \includegraphics[width=.24\linewidth,clip,keepaspectratio]{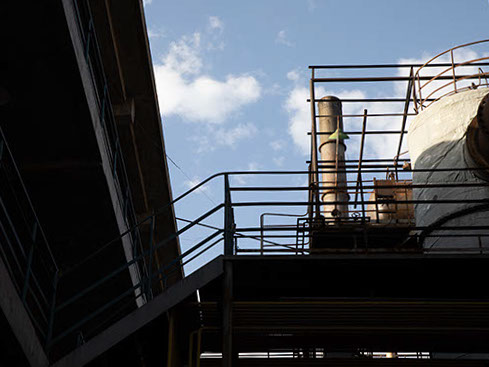} & \includegraphics[width=.24\linewidth,clip,keepaspectratio]{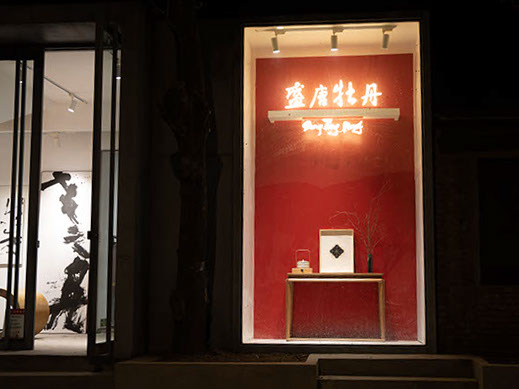} \\
		\rotatebox{90}{\quad\;\;3EV} &
		\includegraphics[width=.24\linewidth,clip,keepaspectratio]{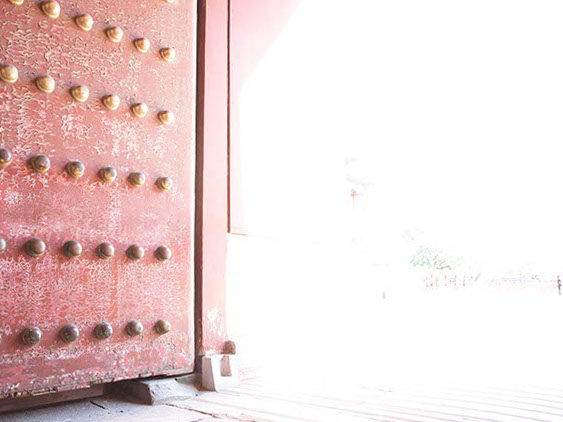} & \includegraphics[width=.24\linewidth,clip,keepaspectratio]{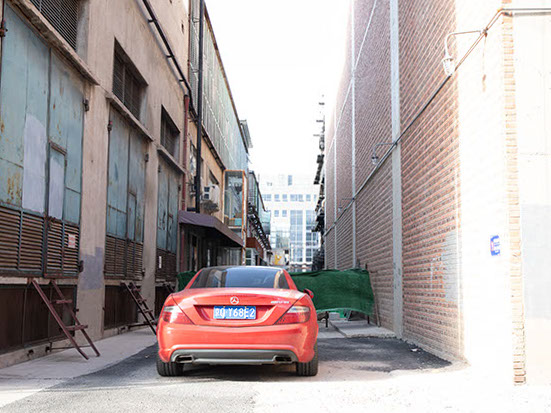} & \includegraphics[width=.24\linewidth,clip,keepaspectratio]{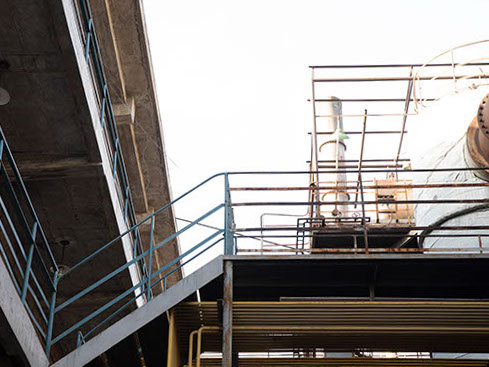} & \includegraphics[width=.24\linewidth,clip,keepaspectratio]{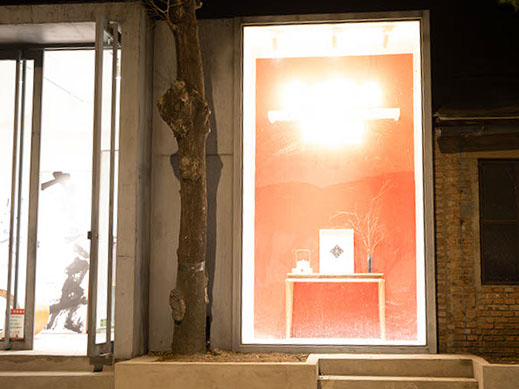} \\
		\rotatebox{90}{\quad \;\,HDR} &
		\includegraphics[width=.24\linewidth,clip,keepaspectratio]{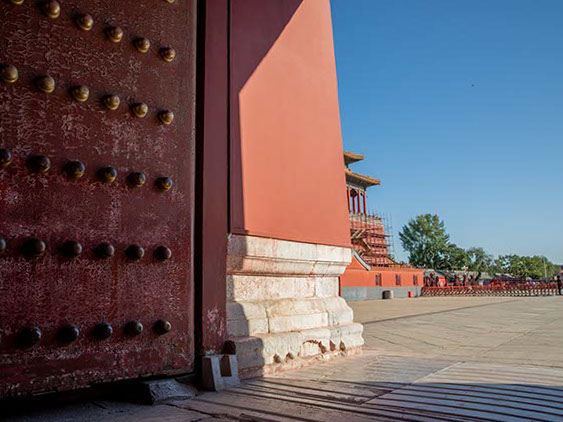} & \includegraphics[width=.24\linewidth,clip,keepaspectratio]{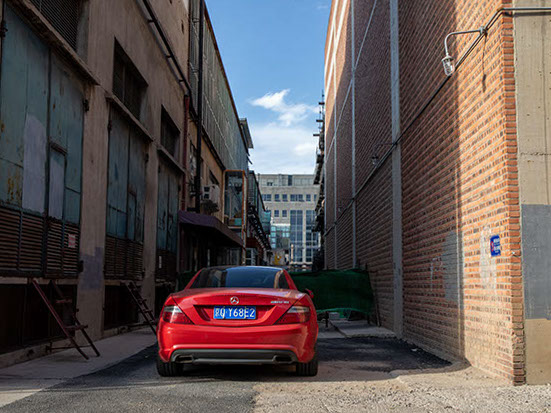} & \includegraphics[width=.24\linewidth,clip,keepaspectratio]{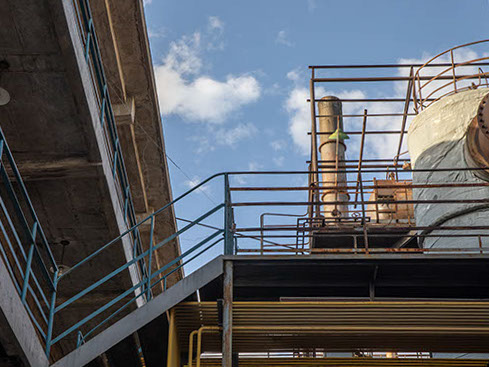} & \includegraphics[width=.24\linewidth,clip,keepaspectratio]{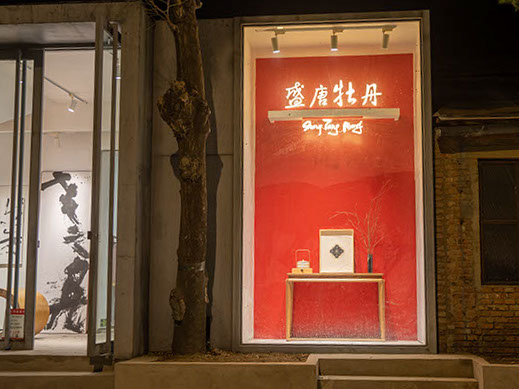}
	\end{tabular}
	\vspace{-2mm}
	\caption{Representative scenes of our captured Raw-to-HDR dataset.}
	\label{fig:dataset}
	\vspace{-3mm}
\end{figure}

\section{Experiments}
In this section, we first introduce the experimental settings, including the compared methods and the evaluation metrics. Then, we conduct experiments on our Raw-to-HDR dataset. Next, we perform ablation studies, to validate the superiority of the proposed model. Finally, we perform extended experiments for further evaluation.

\subsection{Experimental Settings}
\noindent\textbf{Compared methods}. To assess the performance of our RawHDR model, we compare with leading single-exposure HDR reconstruction methods, including HDRCNN~\cite{eilertsen2017hdr}, ExpandNet~\cite{marnerides2018expandnet}, DeepHDR~\cite{Marcel:2020:LDRHDR}, and HDRUNet~\cite{chen2021hdrunet}. In addition, we also compare with methods that are designed to deal with under-exposed regions (SID~\cite{chen2018learning}) and over-exposed regions (EC~\cite{afifi2021learning}). Besides our original model, we incorporate a small version (denoted as Ours-S) by simply reducing the number of layers and channel numbers. 

\vspace{1mm}
\noindent\textbf{Evaluation metrics}. Peak signal-to-noise ratio (PSNR) is a widely used metric to evaluate pixel-level fidelity. We compute PSNR directly on HDR images in linear space. Since HDR images generally need to be tone-mapped before displaying on modern displays, we also test the PSNR on the tone-mapped HDR images. The results are denoted as PSNR-$\mu$, where $\mu$ is the tone mapping parameter. Compared with PSNR, which may stress much importance on the brightest regions of the image, PSNR-$\mu$ is a more balanced metric that can represent the reconstruction fidelity in both dark and bright areas. In other words, PSNR-$\mu$ reveals more information from the HDR display perspective. In addition to pixel-level evaluations, we also employ Structural Similarity~\cite{wang2004image} (SSIM) and multi-scale SSIM~\cite{wang2003multiscale} (MS-SSIM) to evaluate the structural  similarity of the tone-mapped images. Larger PSNR, PSNR-$\mu$, SSIM, and  MS-SSIM show better performances. For each method, we also report the computational cost (GMACs) and number of parameters to evaluate the mode size.

\subsection{Results on Our Raw-to-HDR Dataset}

\begin{table*}[t]
	\centering
	\setlength{\tabcolsep}{2.5mm}
	\footnotesize
	
		\vspace{-2mm}
		\caption{The comparisons of HDR reconstruction performances on our dataset. The best results are highlighted in \textbf{bold}.}
		\vspace{-2mm}
			\begin{tabular}{c|cccccccc}
				\hline\hline
				Metrics &HDRCNN~\cite{eilertsen2017hdr} &ExpandNet~\cite{marnerides2018expandnet}&SID~\cite{chen2018learning}& DeepHDR~\cite{Marcel:2020:LDRHDR}&EC~\cite{afifi2021learning}& HDRUNet~\cite{chen2021hdrunet} &Ours-S & Ours\\
				\hline
				MACs(G) & 54.46 & 13.75 & 13.73 & 19.13 & \textbf{10.57} & 23.61 & 17.85 & 42.07\\
				\hline
				Params(M) & 155.45 & \textbf{0.485} & 7.76 & 51.55 & 7.02 & 1.65 & 4.47 & 10.49 \\
				\hline
				PSNR & 28.13&36.18 & 36.76 & 36.77& 35.16 & 36.67 & 37.03 & \textbf{37.24}\\
				\hline
				PSNR-$\mu$ &18.04& 41.15&37.86 & 41.18&36.15 & 41.16 & 41.48 & \textbf{41.95} \\
				\hline
				SSIM & 0.4360&0.9701&0.9689 & \textbf{0.9739} & 0.9423 &  0.9232& 0.9716 & 0.9714  \\
				\hline
				MS-SSIM& 0.7446 & 0.9919&0.9925 & 0.9910 &0.9806& 0.9851& 0.9921& \textbf{0.9934}\\
				\hline\hline
			\end{tabular}
			\label{tab:real}
	\end{table*}

		\begin{figure*}[t] \small
			\centering
			\setlength{\tabcolsep}{1pt}
			\begin{tabular}{cccccccc}
				
				\includegraphics[width=0.12\linewidth]{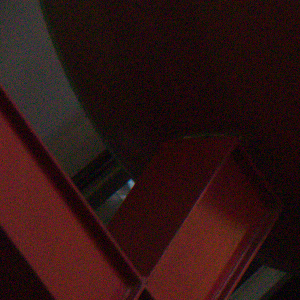} 
				& \includegraphics[width=0.12\linewidth]{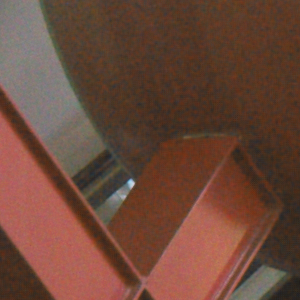}
				& \includegraphics[width=0.12\linewidth]{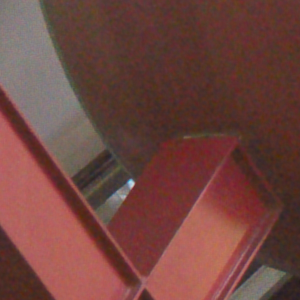}
				&\includegraphics[width=0.12\linewidth]{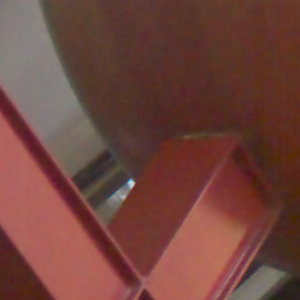}
				&\includegraphics[width=0.12\linewidth]{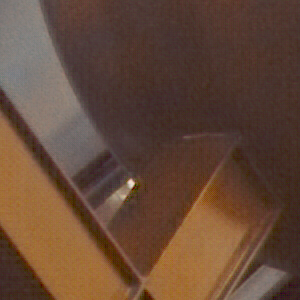}
				&\includegraphics[width=0.12\linewidth]{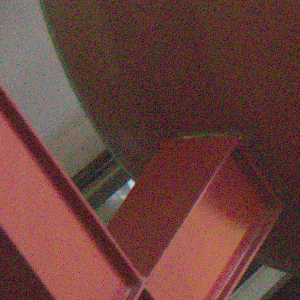}
				&\includegraphics[width=0.12\linewidth]{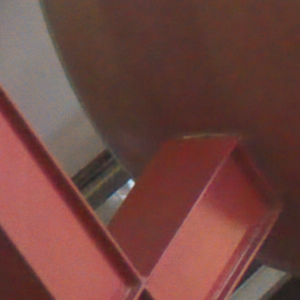}
				&\includegraphics[width=0.12\linewidth]{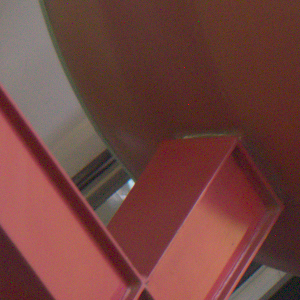}\\
				\includegraphics[width=0.12\linewidth]{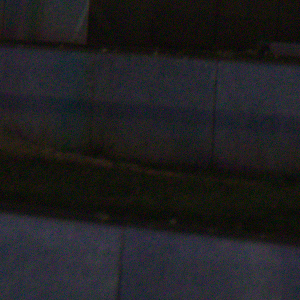} 
				& \includegraphics[width=0.12\linewidth]{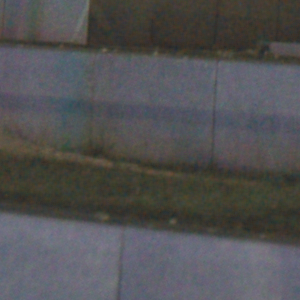}
				& \includegraphics[width=0.12\linewidth]{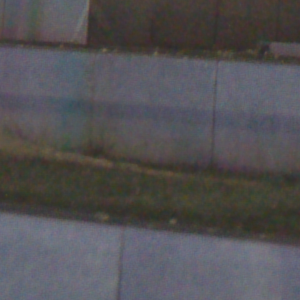}
				&\includegraphics[width=0.12\linewidth]{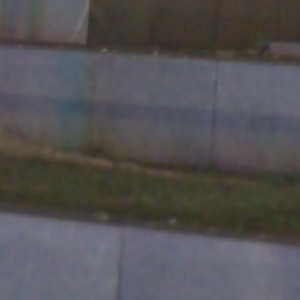}
				&\includegraphics[width=0.12\linewidth]{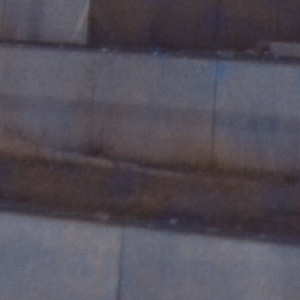}
				&\includegraphics[width=0.12\linewidth]{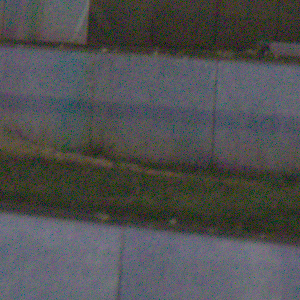}
				&\includegraphics[width=0.12\linewidth]{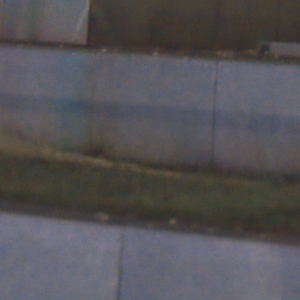}
				&\includegraphics[width=0.12\linewidth]{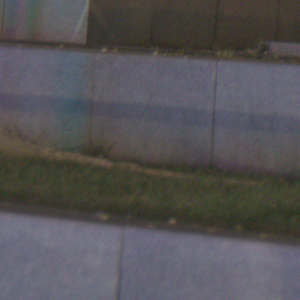}\\
				Input  & ExpandNet~\cite{marnerides2018expandnet} &  SID~\cite{chen2018learning} & DeepHDR~\cite{Marcel:2020:LDRHDR} & EC~\cite{afifi2021learning} &  HDRUNet~\cite{chen2021hdrunet} &  Ours  &  GT \\

			\end{tabular}
		\vspace{-2mm}
			\caption{The result on our dataset. The visualization of typical \textit{under-exposed} regions of Input/ExpandNet/SID/ DeepHDR/EC/HDRUNet/Ours/GT are presented. All images except the input are visualized through tone mapping.}
			\label{fig:dark}
		\end{figure*}
		
		\begin{figure*}[t] \small
			\centering
			\setlength{\tabcolsep}{1pt}
			\begin{tabular}{cccccccc}
				\includegraphics[width=0.12\linewidth]{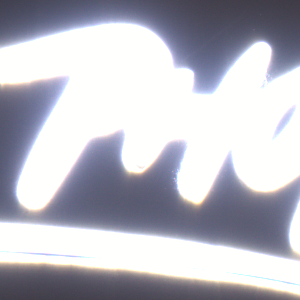} 
				& \includegraphics[width=0.12\linewidth]{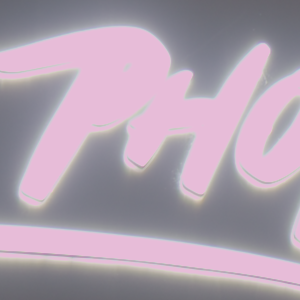}				& \includegraphics[width=0.12\linewidth]{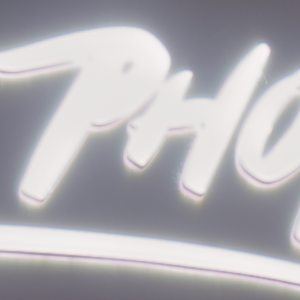}
				&\includegraphics[width=0.12\linewidth]{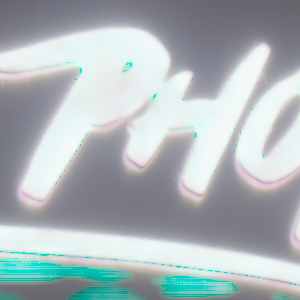}
				&\includegraphics[width=0.12\linewidth]{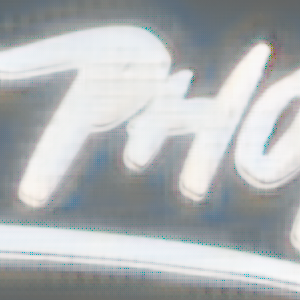}
				&\includegraphics[width=0.12\linewidth]{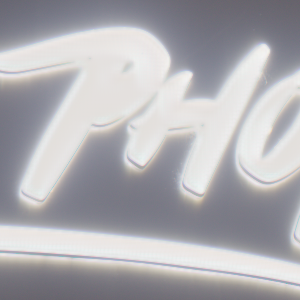}
				&\includegraphics[width=0.12\linewidth]{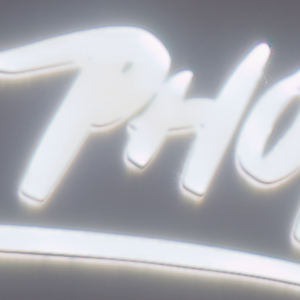}
				&\includegraphics[width=0.12\linewidth]{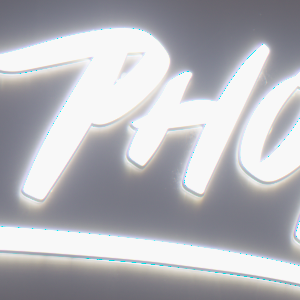}\\
				\includegraphics[width=0.12\linewidth]{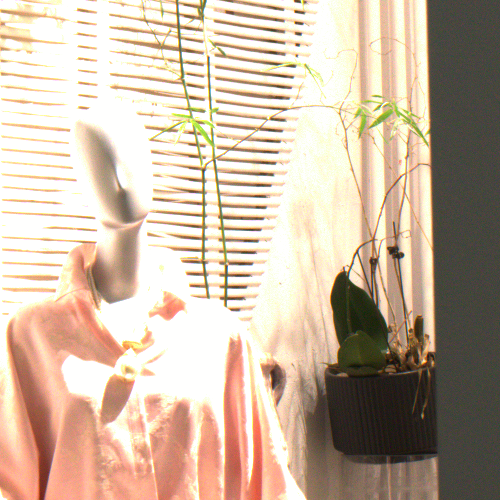} 
				& \includegraphics[width=0.12\linewidth]{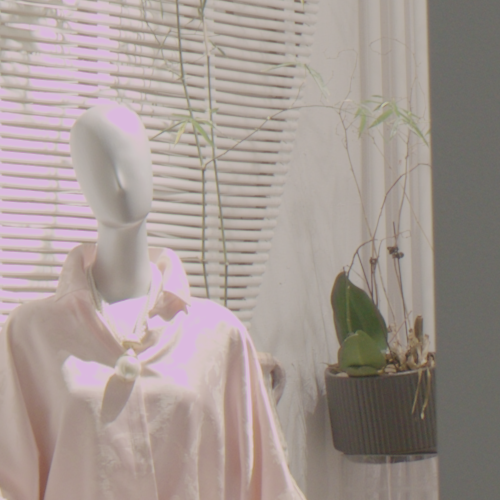}
				& \includegraphics[width=0.12\linewidth]{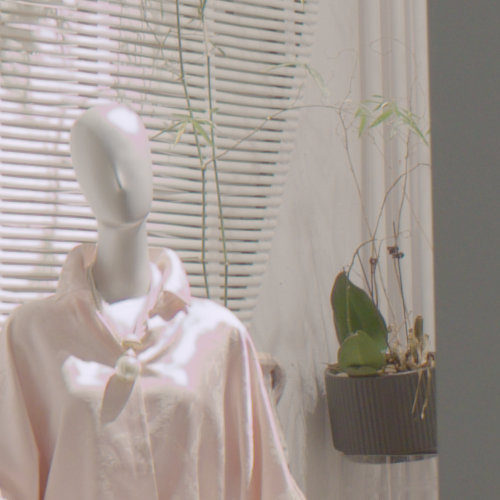}
				&\includegraphics[width=0.12\linewidth]{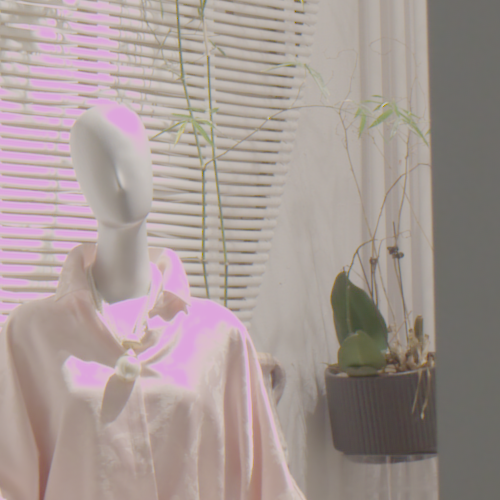}
				&\includegraphics[width=0.12\linewidth]{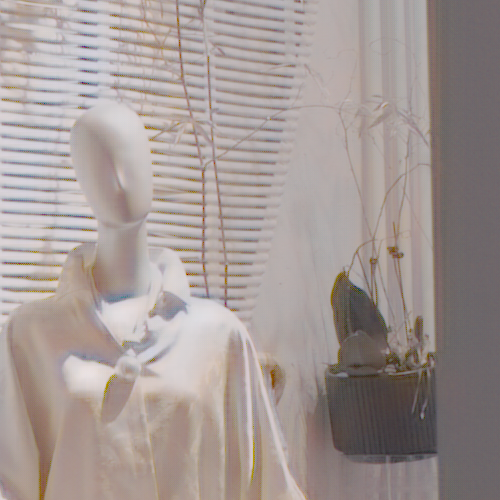}
				&\includegraphics[width=0.12\linewidth]{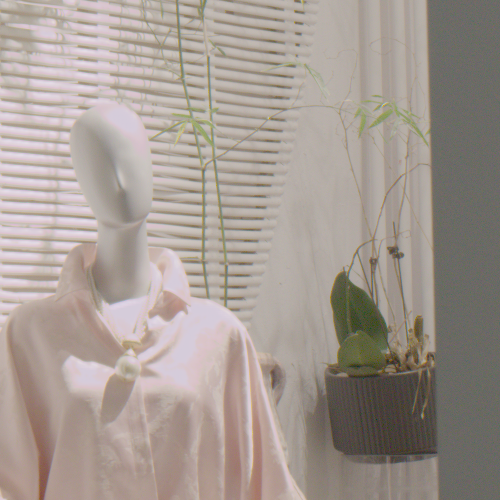}
				&\includegraphics[width=0.12\linewidth]{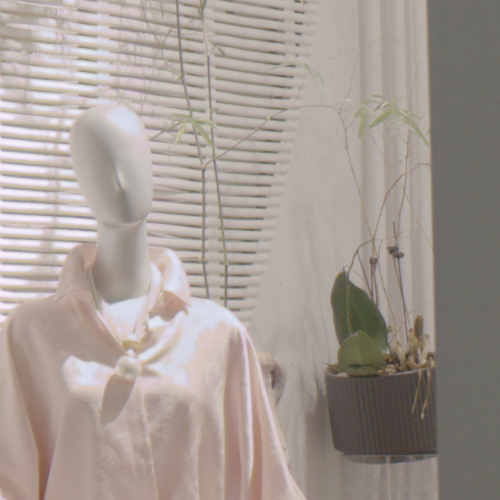}
				&\includegraphics[width=0.12\linewidth]{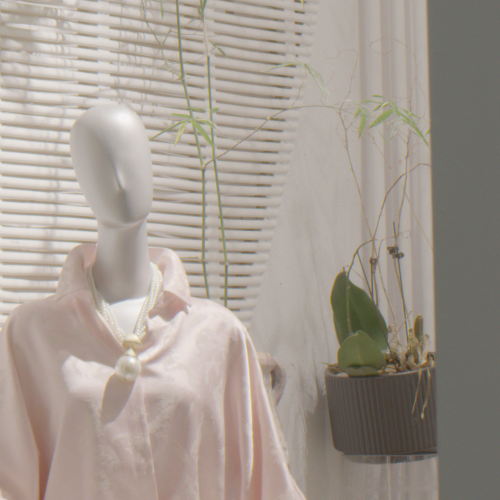}\\
				Input  & ExpandNet~\cite{marnerides2018expandnet} &  SID~\cite{chen2018learning} & DeepHDR~\cite{Marcel:2020:LDRHDR} & EC~\cite{afifi2021learning} &  HDRUNet~\cite{chen2021hdrunet} &  Ours  &  GT \\ 
				
			\end{tabular}
			\caption{The result on our dataset. The visualizations of typical \textit{over-exposed} regions of Input/ExpandNet/SID/ DeepHDR/EC/HDRUNet/Ours/GT are presented. All images except the input are visualized through tone mapping.}
			\label{fig:bright}
		\end{figure*}
		
		To verify the effectiveness of our method, we compare our RawHDR model against prevailing methods on our Raw-to-HDR dataset. Notably, the original architectures of EC~\cite{afifi2021learning}, HDRCNN~\cite{eilertsen2017hdr}, and HDRUNet~\cite{chen2021hdrunet} are designed to accommodate 3-channel RGB images. We adapted these models by modifying their input to 4 channels, ensuring compatibility with the Raw-to-HDR framework. Table~\ref{tab:real} summarizes the numerical results according to the averaged values of all evaluation metrics. We can see that our method outperforms competing methods in PSNR, PSNR-$\mu$, MS-SSIM, and presents a competitive performance in SSIM. One observation is that the advantage of our RawHDR is greater in PSNR-$\mu$, compared with PSNR. This is benefited from our specially designed deep architecture for hard region recovery. The qualitative results are shown in Figures~\ref{fig:dark}-\ref{fig:bright}. It can be seen that the images recovered from our method approximate ground truth well and is significantly better than other reconstruction methods, especially for extremely dark and bright regions. This phenomenon is benefited from our dual intensity and global spatial guidance, which better exploits features from Raw images.

		\subsection{Ablation Studies}
		
		This section performs ablation studies to verify the effectiveness of our model components and the proposed dataset.
		
\begin{figure*}[t]
			\centering
			\begin{subfigure}[b]{.12\linewidth}
				\centering
				\includegraphics[width=\linewidth]{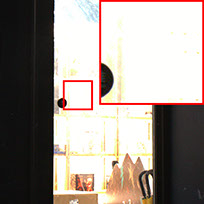}
				\caption{Input}
			\end{subfigure}
			\begin{subfigure}[b]{.12\linewidth}
				\centering
				\includegraphics[width=\linewidth]{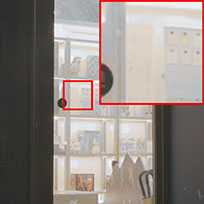}
				\caption{Hard mask}
				\label{fig:hard}
			\end{subfigure}
			\begin{subfigure}[b]{.12\linewidth}
				\centering
				\includegraphics[width=\linewidth]{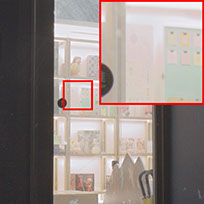}
				\caption{w/o GSG}
				\label{fig:wogsg}
			\end{subfigure}
			\begin{subfigure}[b]{.12\linewidth}
				\centering
				\includegraphics[width=\linewidth]{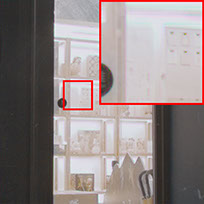}
				\caption{w/o DIG}
				\label{fig:wodig}
			\end{subfigure}
			\begin{subfigure}[b]{.12\linewidth}
				\centering
				\includegraphics[width=\linewidth]{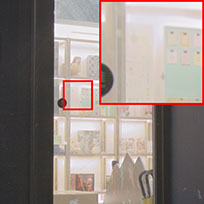}
				\caption{G guid.}
			\end{subfigure}
			\begin{subfigure}[b]{.12\linewidth}
				\centering
				\includegraphics[width=\linewidth]{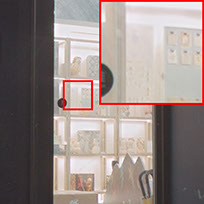}
				\caption{RB guid.}
				\label{fig:rb}
			\end{subfigure}
			\begin{subfigure}[b]{.12\linewidth}
				\centering
				\includegraphics[width=\linewidth]{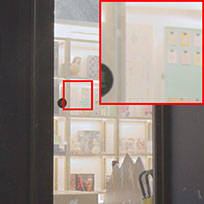}
				\caption{Ours}
			\end{subfigure}
			\begin{subfigure}[b]{.12\linewidth}
				\centering
				\includegraphics[width=\linewidth]{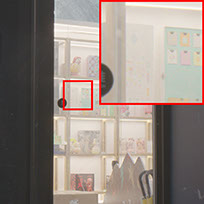}
				\caption{GT}
			\end{subfigure}
			\vspace{-3mm}
			\caption{The visualization results for ablation studies.}
			\vspace{-3mm}
			\label{fig:ablation}
			
		\end{figure*}

		\vspace{1mm}
		\noindent\textbf{Mask estimation.}
		First, we evaluate the effectiveness of the proposed mask estimation module. In this module, we use deep neural networks to learn a soft mask, which well separates the over- and under-exposed areas. Here, we also conduct ablation studies that uses hard masks  $M_{over}^h$ and $M_{under}^h$.  From the results in Figure~\ref{fig:ablation}\sref{fig:hard}, we see that using hard masks suffers from color artifacts, which do not appear in the results of our soft mask estimation.
		
		\vspace{1mm}
		\noindent\textbf{Dual intensity guidance.} Here, we evaluate the promotion brought by the dual intensity guidance. As shown in Table~\ref{tab:ablation}, by replacing each or both channel intensity guidance (denoted as w/o DIG), we notice apparent performance degradation. From the visual results illustrated in Figure~\ref{fig:ablation}\sref{fig:wogsg}, we see more details are recovered in hard regions with the help of the dual intensity guidance.
		
		\vspace{1mm}
		\noindent\textbf{Global spatial guidance.} As shown in Table~\ref{tab:ablation} and Figures~\ref{fig:ablation}\sref{fig:wodig}-\sref{fig:rb}, we also evaluate our global spatial guidance by removing this module (denoted as w/o GSG). By comparing both quantitative and qualitative results, we can infer that global spatial guidance indeed plays an important role in extremely dark and bright regions.
		
		\begin{table}[t]
			\centering
			\setlength{\tabcolsep}{3mm}
			\setlength{\abovecaptionskip}{2mm} 
			\footnotesize
			\begin{threeparttable}
				\caption{The ablation study results on network design.}
				\begin{tabular}{c|cccc}
					\hline\hline
					Setting & PSNR  & PSNR-$\mu$ &  SSIM & MS-SSIM \\
					\hline
					Hard mask &35.70&39.37&0.9617&0.9904\\
					\hline
					w/o DIG & 35.41 &38.27 &  0.9206 & 0.9826\\
					\hline
					G guid. & 35.85 & 40.30 &0.9643 & 0.9913 \\
					\hline
					RB guid. &36.59&38.72& 0.9561 & 0.9889 \\
					\hline
					w/o GSG &36.45&41.04&0.9687&0.9933\\
					\hline
					Ours&\textbf{37.04}&\textbf{41.50}&\textbf{0.9707}&\textbf{0.9935}\\
					\hline\hline
					
				\end{tabular}
				\label{tab:ablation}
			\end{threeparttable}
			\vspace{-2mm}
		\end{table}
		
		\subsection{Further Evaluation for Raw-to-HDR}
		In this section, we further conduct extended experiments to verify the effectiveness of the setting to reconstruction HDR from Raw images.
		
\begin{figure}[t]
			\centering
			\begin{subfigure}[b]{.24\linewidth}
				\centering
				\includegraphics[width=\linewidth]{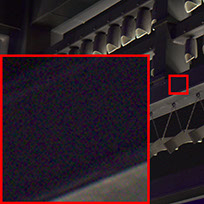}
				\subcaption*{Input}
			\end{subfigure}
			\begin{subfigure}[b]{.24\linewidth}
				\centering
				\includegraphics[width=\linewidth]{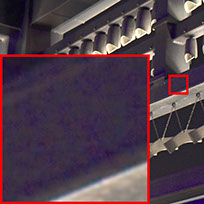}
				\subcaption*{Ours (RGB)}
			\end{subfigure}
			\begin{subfigure}[b]{.24\linewidth}
				\centering
				\includegraphics[width=\linewidth]{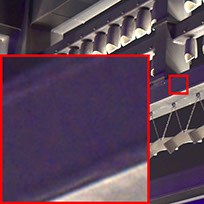}
				\subcaption*{Ours (Raw)}
			\end{subfigure}
			\begin{subfigure}[b]{.24\linewidth}
				\centering
				\includegraphics[width=\linewidth]{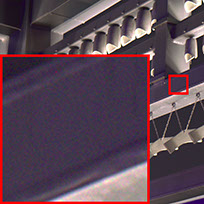}
				\subcaption*{GT}
			\end{subfigure}
			\vspace{-2mm}
			\caption{The quantitative results for our RawHDR model trained on sRGB data and Raw data.}
			\vspace{-2mm}
			\label{fig:rgb}
			
		\end{figure}

		\vspace{1mm}
		\noindent\textbf{Evaluation of the Raw-to-HDR mapping.} 
		To validate the superiority of Raw-to-HDR reconstruction compared with commonly used sRGB-to-HDR, we design experiments to compare sRGB and Raw data for HDR reconstruction. Specifically, we synthesize sRGB images from the Raw data of the proposed dataset through a simple camera ISP (following~\cite{abdelhamed2018high}), and then form paired sRGB/HDR dataset. We train our RawHDR model on both the newly curated sRGB/HDR dataset and the proposed dataset in the same setting. For the results of Raw-to-HDR setting, we use the same processing pipeline to guarantee that both settings are in the same color space. Results in Table~\ref{tab:rgb} and Figure~\ref{fig:rgb} validate that by introducing linear high-bit Raw images, the reconstruction of HDR suffers less from information loss. Therefore, in order to obtain high-quality HDR, it is necessary to feed HDR reconstruction model with Raw images.
		
		\begin{table}[t]
			\centering
			\setlength{\tabcolsep}{3mm}
			\setlength{\abovecaptionskip}{1.5mm} 
			\footnotesize
			\begin{threeparttable}
				\caption{The comparison between sRGB and Raw.}
				\begin{tabular}{c|cccc}
					\hline\hline
					Setting & PSNR  & PSNR-$\mu$ &  SSIM & MS-SSIM \\
					\hline
					sRGB & 37.25 & 37.82 & 0.9385 & 0.9798\\
					\hline
					Raw&\textbf{44.15}&\textbf{39.76}&\textbf{0.9509}&\textbf{0.9848}\\
					\hline\hline
				\end{tabular}
				\label{tab:rgb}
			\end{threeparttable}
			\vspace{-3mm}
		\end{table}
	
\begin{figure}[t]
			\centering
			\begin{subfigure}[b]{.24\linewidth}
				\centering
				\includegraphics[width=\linewidth]{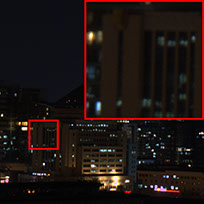}
				\subcaption*{Input}
			\end{subfigure}
			\begin{subfigure}[b]{.24\linewidth}
				\centering
				\includegraphics[width=\linewidth]{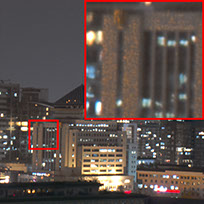}
				\subcaption*{Ours (RGB)}
			\end{subfigure}
			\begin{subfigure}[b]{.24\linewidth}
				\centering
				\includegraphics[width=\linewidth]{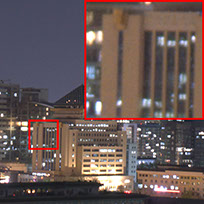}
				\subcaption*{Ours (Raw)}
			\end{subfigure}
			\begin{subfigure}[b]{.24\linewidth}
				\centering
				\includegraphics[width=\linewidth]{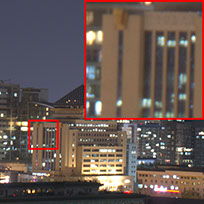}
				\subcaption*{GT}
			\end{subfigure}
			\vspace{-2.5mm}
			\caption{The results for cross-camera setting. The model is trained on Canon 5D IV and evaluated on Sony A7R4.}
			\vspace{-5mm}
			\label{fig:othercamera}
			
		\end{figure}

		\vspace{1mm}
		\noindent\textbf{Cross-camera HDR reconstruction}.
		Besides the advantage that Raw images have higher bit-depth than sRGB (14-bit \vs 8-bit), another crucial attribute is the linearity.  When attempting HDR reconstruction from a singular sRGB image, the learned models tend to be camera-specific, since the network is overfitted to a specific camera processing pipeline. However, Raw images are in linear space and unprocessed by any in-camera operation. As a result, discrepancies between Raw images from various sources are much less noticeable than those in sRGB format. To validate this, we capture Raw images from a different camera, \ie, Sony A7R4, and directly evaluate the HDR reconstruction results using the model trained on Canon 5D IV. As shown in Figure~\ref{fig:othercamera}, when transferring to another camera, the model trained in sRGB setting fails to produce high-quality HDR, while the results are robust using Raw data.
		
		\section{Conclustion}
		\vspace{-2mm}
		In this work, we aim to solve the most challenging problem in single exposure HDR reconstruction, which is the insufficient information in the darkest and brightest regions of a high dynamic scene. Our work solves this problem from two aspects. First, we propose to reconstruct HDR from a single Raw image instead of sRGB, since Raw images have higher bit-depth and intensity tolerance. Then, we propose a novel deep model, \ie, RawHDR, to learn the Raw-to-HDR mapping. Specifically, the proposed mask estimation, dual intensity guidance, and global spatial guidance help us to separate hard and easy regions and fully exploit local and long range features. We capture high-quality Raw-to-HDR datasets for data-driven methods, and our dataset is evaluated to be very necessary for HDR reconstruction. Experimental results verify that both our dataset and RawHDR model are of high quality. The Raw-to-HDR mapping can be potentially integrated in real camera ISP once the model is simplified, and we remain this as our future work.
		
		\vspace{1mm}
		\noindent\textbf{Acknowledgments}
		This work was supported by the National Natural Science Foundation of China (62171038, 61931008, 62088101, and U21B2024), the R\&D Program of Beijing Municipal Education Commission (KZ202211417048), and the Fundamental Research Funds for the Central Universities.

{\small
\bibliographystyle{ieee_fullname}
\bibliography{egbib}
}

\end{document}